\documentclass{vgtc}                          

\ifpdf
  \pdfoutput=1\relax                   
  \usepackage{graphicx}                
  \DeclareGraphicsExtensions{.pdf,.png,.jpg,.jpeg} 
\else
  \usepackage{graphicx}                
  \DeclareGraphicsExtensions{.eps}     
\fi%

\graphicspath{{figures/}{pictures/}{images/}{./}} 

\usepackage{microtype}                 
\PassOptionsToPackage{warn}{textcomp}  
\usepackage{textcomp}                  
\usepackage{mathptmx}                  
\usepackage{times}                     
\usepackage{cite}                      
\usepackage{tabu}                      
\usepackage{booktabs}                  

\onlineid{1023}

\vgtccategory{Technical Papers}

\vgtcinsertpkg

\usepackage{tikz}
\newcommand\copyrighttext{%
  \tiny \textcopyright 2019 IEEE. Personal use of this material is permitted. Permission from IEEE must be obtained for all other uses, in any current or future media, including reprinting/republishing this material for advertising or promotional purposes, creating new collective works, for resale or redistribution to servers or lists, or reuse of any copyrighted component of this work in other works. Cite this article as follows: R. O\v{s}lej\v{s}ek, V. Rus\v{n}\'{a}k, K. Bursk\'{a}, V. \v{S}v\'{a}bensk\'{y} and J. Vykopal, "Visual Feedback for Players of Multi-Level Capture the Flag Games: Field Usability Study," \textit{2019 IEEE Symposium on Visualization for Cyber Security (VizSec)}, Vancouver, BC, Canada, 2019, pp. 1-11, doi: \url{https://doi.org/10.1109/VizSec48167.2019.9161386}.}
\newcommand\copyrightnotice{%
\begin{tikzpicture}[remember picture,overlay]
\node[anchor=south,yshift=12pt] at (current page.south) {\fbox{\parbox{\dimexpr\textwidth-\fboxsep-\fboxrule\relax}{\copyrighttext}}};
\end{tikzpicture}%
}

\usepackage{multirow}
\usepackage{pifont}
\usepackage{makecell}
\usepackage{booktabs}

\title{Visual Feedback for Players of Multi-Level Capture the Flag Games:\\ Field Usability Study} 

\author{Radek O\v{s}lej\v{s}ek\thanks{e-mail: oslejsek@fi.muni.cz}\\ %
        \parbox{1.5in}{\scriptsize \centering Masaryk University\\ Faculty of Informatics} %
\and V\'{i}t Rus\v{n}\'{a}k\thanks{e-mail: rusnak@ics.muni.cz}\\ %
        \parbox{1.5in}{\scriptsize \centering Masaryk University\\ Institute of Computer Science} %
\and Karol\'{i}na Bursk\'{a}\thanks{e-mail: burska@fi.muni.cz}\\ %
        \parbox{1.5in}{\scriptsize \centering Masaryk University\\ Faculty of Informatics} %
\and Valdemar \v{S}v\'{a}bensk\'{y}\thanks{e-mail: svabensky@ics.muni.cz}\\ %
        \parbox{1.5in}{\scriptsize \centering Masaryk University\\ Faculty of Informatics} %
\and Jan Vykopal \thanks{e-mail: vykopal@ics.muni.cz} \\ %
        \parbox{1.5in}{\scriptsize \centering Masaryk University\\ Institute of Computer Science}} %

\teaser{
  \includegraphics[width=\linewidth]{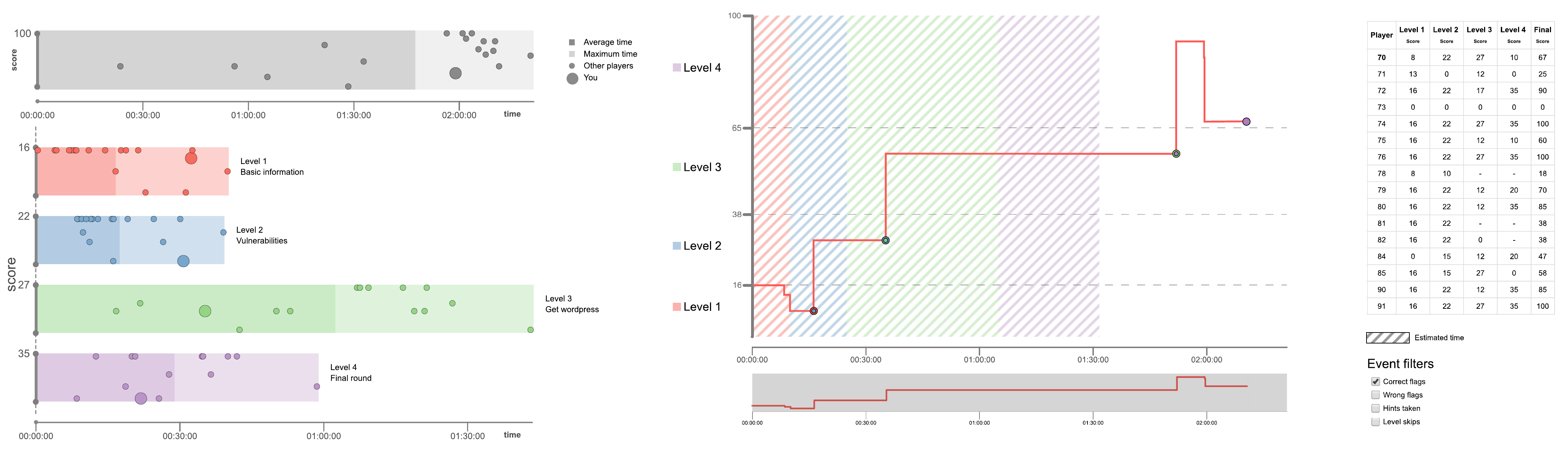}
  \centering
   \caption{\textsc{Clustering} (left) and \textsc{Timeline} (right) interactive visualizations provide feedback to players of serious multi-level cybersecurity games. They offer holistic as well as person-centric perspectives on the score development of one or more players.}
 \label{fig:teaser}
}

\abstract{Capture the Flag games represent a popular method of cybersecurity training. Providing meaningful insight into the training progress is essential for increasing learning impact and supporting participants' motivation, especially in advanced hands-on courses. In this paper, we investigate how to provide valuable post-game feedback to players of serious cybersecurity games through interactive visualizations. In collaboration with domain experts, we formulated user requirements that cover three cognitive perspectives: gameplay overview, person-centric view, and comparative feedback. Based on these requirements, we designed two interactive visualizations that provide complementary views on game results. They combine a known clustering and time-based visual approaches to show game results in a way that is easy to decode for players. The purposefulness of our visual feedback was evaluated in a usability field study with attendees of the Summer School in Cyber Security. The evaluation confirmed the adequacy of the two visualizations for instant post-game feedback. Despite our initial expectations, there was no strong preference for neither of the visualizations in solving different tasks.
}


\CCScatlist{
  \CCScatTwelve{Human-centered computing}{Visual analytics}{}{};
  \CCScatTwelve{Human-centered computing}{Empirical studies in interaction design}{}{};
  \CCScatTwelve{Human-centered computing}{Visualization design and evaluation methods}{}{};
  \CCScatTwelve{Applied computing}{Interactive learning environments}{}{}
}

\begin{document}

\firstsection{Introduction} 

\maketitle
\copyrightnotice

As cyber attacks have been on the rise in recent years, security professionals and students have to be trained in adversary thinking, which enables them to understand cyber attacks and set up effective defenses. A popular way of cybersecurity training is through gamification and serious games~\cite{abt1987,michael2005}. A general shortage of methodologies and tools for timely feedback in the field of serious games is emphasized in~\cite{arnab15, bellotti13}. This deficiency is even more apparent for cybersecurity serious games, which pose specific demands on environment capabilities~\cite{stewart09}.

The subject of our research is to provide meaningful insight into the training progress. In this paper, we investigate how to provide valuable visual feedback to players of serious cybersecurity games right after the exercise so that they can immediately learn from their experience and compare their results with other players. Following the terminology used in the cybersecurity training domain, players of serious games are referred to as \emph{trainees} in this paper.

Serious games are of many types. To reach the goal, we restrict ourselves to Capture the Flag (CTF) games~\cite{Werther2011,boopathi2015,svabensky2018} that are played in virtual environments, in which gameplay events can be monitored and used for providing automated visual feedback to trainees. 

This paper deals with a multi-level variant of CTF games. Each game, regardless of its specific objectives or content, consists of well-described tasks divided into consecutive levels. The access to the next level is conditioned by fulfilling tasks from the previous one. Moreover, players can take hints or skip the entire level. Points are awarded or deducted for these actions so that the final scores of individual players are mutually comparable and can be used for their evaluation. 

The design and development of our feedback visualizations went through four stages: (1) understanding of serious multi-level games, their objectives in cybersecurity domain, and available data, (2) defining requirements on the visual feedback in accord to the educational goals of the games, (3) prototyping, iterative design and development, and (4) usability study performed at our university and involving students of our cybersecurity training lessons.

\textbf{Contributions.} The main contributions of the paper are: (a) we classified the visual feedback requirements into three categories that cover trainees' expectations of the training (personalized feedback, comparative feedback, and overall results); (b) we applied existing visualization techniques in the domain of hands-on cybersecurity training in order to provide better insight into the trainees' results right after the training session; (c) we performed a formal evaluation that confirmed the meaningfulness of the defined requirements and usefulness of the post-game analysis visualizations for trainees.

The remainder of the paper is organized as follows: Section~\ref{sec:related-work} introduces the related work in the area of visual analysis of serious games, particularly in the cybersecurity domain. Section~\ref{sec:dataset} describes available data and provides the example of a cybersecurity game. In Section~\ref{sec:requirements}, we formulate requirements posed on visualizations and corresponding tasks covering three different cognitive perspectives on game results. We discuss our approach to fast visual feedback in Section~\ref{sec:visual-design}. Section~\ref{sec:method} describes the usability study and brings necessary details about the usability testing concerning defined hypotheses. Results of the usability study are discussed in Section~\ref{sec:discussion}. We draw our conclusions and look to the future in Section~\ref{sec:conclusion}.

\section{Related Work}
\label{sec:related-work}

Many works published in the field of user behavior visual analysis focus on social media. Cao et al.~\cite{Cao2016}, for instance, proposed a visual analysis tool for anomalous user behavior in online communication systems and social media platforms. The proposed system incorporates presentation of user's communication activities and interactions. Another analysis-driven approach to user behavior was introduced by Kumar et al.~\cite{kumar2011}. They investigated users’ migration behaviors among various social media platforms and represented the findings via radar charts. Many other works also address the topic of social media-related user behavior in varied scopes, such as ~\cite{Yang2010,thom2012,cao2015}. A comprehensive survey on visual approaches to the analysis of anomalous users can be found in~\cite{shi2019}. Our solution also supports behavioral analysis but in a very simplified way enabling players of CTF games to learn from their behavior by being aware of their steps and steps of other players.

The positive effects of visual analytics integration into the learning process have already been identified. The outcomes serve for understanding trainees' actions or optimization of the learning environment~\cite{Vatrapu11, loh15}.
As the educational visualization dashboards have gained considerable attention in the field of learning analytics, reviews concerning this matter emerged as well~\cite{bodily2017,schwendimann2017}.
The visualizations can monitor trainee's progress and help to compare the performance with other peers~\cite{govaerts12}. They can also increase motivation and encourage trainees to compete or to collaborate~\cite{govaerts10}. In this paper, we focus on automatically generated post-training feedback. In~\cite{Corrin2015}, the authors investigated how students interpret feedback delivered via learning analytics dashboards in distinct courses. Their findings reveal that the majority (83\%) of students were able to identify gaps in their performance.

Various general tools for qualitative feedback or assessment in education exist, and many studies include a platform, where the dashboards are employed in multiple field of education. They serve for capturing the behavior of students or provide long term statistics and observations for both students and teachers~\cite{Jivet2018,deFreitas2017,Leony2012,Matcha2019}.
An online tool~\textit{asTTle}, for instance, can assess students' achievements and progress. It allows creating pen-and-paper tests, and then continuously analyzing specified characteristics, which can be stipulated by teachers. The students' outcomes are then visualized in interactive reports that provide rich feedback related to student performance~\cite{hattie06}. \textit{Questionmark Perception}~\cite{questionmark}, another example of a similar system, is used for education in the form of surveys, tests, or exams and enables the creation of reports from these events. Govaerts et al.~\cite{govaerts10, govaerts12} proposed a general-purpose web-based environment for the visualization of the students' progress and results based on the tracking and evaluation of Twitter hashtags. Their tools help the students to assess themselves and to get automated feedback on their achievements throughout an online course.

These works, however, often tend to take a 'one-size-fits-all' approach to the collection, processing, and reporting of data, overlooking disciplinary knowledge practices. Furthermore, since they focus on long-term courses, they are not suitable for our needs as they require different perspectives on the analysis of the learning process. While the useful applications of visual analytics in education are well known, we lack its use in the cybersecurity training. Therefore, as there are learning dashboards designed for teaching specific disciplines like programming~\cite{Grissom2003,Fu2017}, mathematics~\cite{jacobs05}, or specific cyber defense exercises~\cite{vykopal2018}, we aim at providing learning feedback in a specific domain -- CTF games.

A comprehensive list of numerous cybersecurity CTF games can be found at the CTFtime portal~\cite{ctftime.org}. However, most of the listed platforms provide only limited information about collected data and their presentation to the users. 

\emph{CTFd}~\cite{chung2017} is a platform for creating and hosting CTF challenges. It visualizes overall score graphs and breakdowns for individual players. The latter includes percentages of all challenges solved, distributions of solved challenges into categories, and the evolution of player score over time. However, user evaluation of the platform effectiveness is missing.

\emph{EDURange}~\cite{weiss2017magazine} allows gathering the command-line history of a trainee, including command timestamps, their arguments, and exit statuses. The platform can automatically generate an oriented graph that visualizes the command history. The vertices of the graph represent executed commands; the edges represent the sequence of commands (that is, an edge from a command $x$ to $y$ means that $y$ was executed after $x$). Supervisors can use the graphs in real time to check how the trainees progress and whether they need extra guidance. A post-game use case would be to compare the graphs to each other or a previously prepared pattern of a sample solution.

During a \emph{Crossed Swords} exercise~\cite{kont2017}, network traffic, logs, and system activity metrics were collected and analyzed. The goal was to provide real-time feedback and situational awareness for the trainees. In a post-game survey, 11 out of 14 participants found the feedback useful for their learning, and the remaining three were neutral. The authors raise the question of finding a balanced amount of information to provide to participants since 4 of them reported being distracted by the feedback.

\emph{PicoCTF} is an online competition that targets high school students. The competition comprises a series of challenges in the form of an interactive storytelling game~\cite{picoctf}. An evaluation of the game design is based on survey responses and user interaction logs~\cite{picoctf2}. The collected data is not used for assessing the players.

An international \emph{iCTF} competition provides feedback in the form of a scoreboard, which is also used during the game to inform the competitors of the score development and status of their services~\cite{doupe}.

From the information available, most of the CTF platforms offer only a simple scoreboard for comparing players' final score. Techniques used in current security training programs do not facilitate any further summative assessment or feedback for the players regarding their actions in the game~\cite{netwars, nagarajan12}.

\section{Dataset Characteristics and Game Example} 
\label{sec:dataset}

This section describes the data collected during multi-level CTF events. The introduced game example explains available data in detail and demonstrates the principles of cybersecurity CTF games. Moreover, the presented game was selected for our usability study.

\subsection{Selected Cybersecurity Game}

We chose a cybersecurity game named \textit{The Biggest Stock Scam Of All Time}. The game was created by the students of Cyber Attack Simulation course~\cite{ITICSE18kypolab} and was further improved by cybersecurity experts.  
In a background story that complements the game, the trainee takes on a role of a former employee of a global stock trading company. However, he was fired because of refusing to falsify the reports of the company's earnings. When someone else did the job, he wanted to prove the company's corrupt intentions, but to gain evidence, he needs to access the company's records.

\begin{figure}[ht]
  \centering
  \includegraphics[width=0.9\columnwidth]{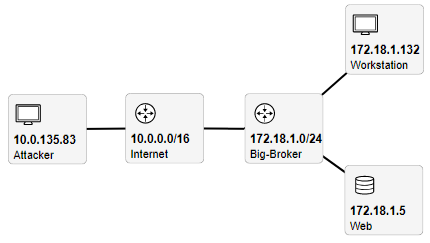}
  \caption{The trainee's view of the game network topology.} 
  \label{fig:game1}
\end{figure}

\begin{figure}[ht]
  \centering
  \includegraphics[width=0.9\columnwidth]{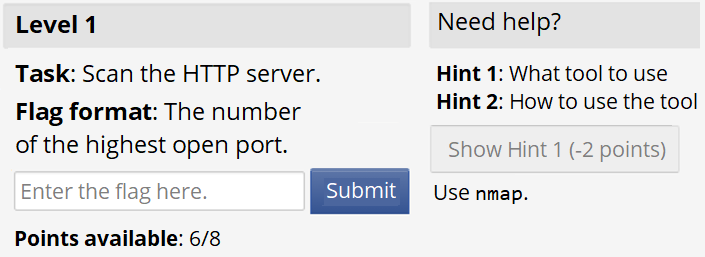}
  \caption{The trainee's view of the game interface.} 
  \label{fig:game2}
\end{figure}

Figure \ref{fig:game1} shows the game's network topology. At the beginning of the game, the trainee receives control of a single attacker virtual machine in a realistic environment emulated by the KYPO cyber range~\cite{KYPO}
The machine runs Kali Linux, 
a standard distribution for penetration testing. The learning objective of the game is to practically exercise cyber attacks. In four consecutive levels, the trainee must gain access to the company's web server, which is available only from an exposed workstation in the company's internal network.
\begin{enumerate}
\item In the first level, the trainee learns about the existence of a company's workstation that is accessible from the internet. An \texttt{nmap} scan reveals that TCP port 22 is open on the workstation, running the SSH service. Having a list of common usernames and passwords, the trainee performs a dictionary attack and accesses the workstation.
\item From the workstation, the trainee can now access the server, which hosts a WordPress website. Since old versions of plugins in WordPress websites are known to contain vulnerabilities, the trainee scans the server using \texttt{wpscan} and reveals a file upload vulnerability.
\item The trainee proceeds to exploit the vulnerability using \emph{Metasploit} penetration testing framework to gain shell access on the server. However, the trainee does not yet possess the necessary privileges to read the file with the stock dealings.
\item Although the server is well-secured, it allows to run \texttt{tcpdump} as a root user, which allows executing scripts. Thus, the trainee escalates the privileges on the file and reads the company's dark secrets.
\end{enumerate}

Figure \ref{fig:game2} shows the game's user interface. Completing each level yields a flag worth a certain number of points that add up to the trainee's score. Each level also contains hints, which the trainee can view in exchange for a scoring penalty. If the trainee is still stuck, (s)he can display the full solution to the level but will not receive any points for it. This scaffolding system allows each trainee to progress individually.

\subsection{Data Logging Explanation}

We record the interaction with the game interface in the form of \textit{game events}. There are eight game events: starting the game, ending the game, starting a level, ending a level by skipping it, ending a level by submitting a correct flag, submitting an incorrect flag, taking a hint, and displaying a solution to the level.

\footnotetext[2]{Supplementary materials also available at \url{https://www.radek-oslejsek.cz/it/supp-materials/}}

Each game event is logged as one line in a CSV file\protect\footnotemark[2]. with the following five-part structure:
\begin{itemize} \itemsep 0pt
    \item \texttt{player\_id} -- a unique numerical ID randomly assigned to each trainee before the game,
    \item \texttt{timestamp} -- absolute time in the format YYYY-MM-DD HH:MM:SS,
    \item \texttt{logical\_time} -- relative time from the start of the level in the format HH:MM:SS,
    \item \texttt{level} -- the order of the level,
    \item \texttt{event} -- one of the eight game events described above.
\end{itemize}
An example of a record in the log is:
\begin{verbatim}
9003581,2018-08-24 16:57:54,00:03:42,4,Hint 3 taken
\end{verbatim}

\section{User Requirements} 
\label{sec:requirements}

Content of CTF games differs. It is therefore impossible to predict and conceptualize content-related questions that trainees would be interested in as part of the post-game feedback. Instead, we focused on capturing high-level user requirements that follow the unified structure of multi-level CTF games and corresponding data. To elicit the requirements, we organized discussions with four domain experts who regularly organize CTF games and who understand educational aspects of training. These experts are skilled in providing informal feedback to players right after the training session and then they have insight into interests of trainees during the post-training debate. Two of the experts are co-authors of this paper. Based on discussion with the experts, we defined three high-level requirements for the visual feedback that helped us to conceptualize views on game data and to design specific visualizations:

\textbf{\textsl{R1:} Provide personalized feedback.} Players should be able to find out their results and identify their well-done and problematic parts of the game. This requirement includes person-centric goals and questions regardless of other players, e.g., ``In which level did I lose the most points and why?''.

\textbf{\textsl{R2:} Provide comparative feedback.} Players should be able to identify parts of the game where they were better or worse than other players. This requirement introduces a competitiveness factor into the feedback, which enables players to compare themselves with others and assess their abilities within a group, typically in a competition.

\textbf{\textsl{R3:} Provide a brief overview of the overall game results and features.} Players should be able to get a necessary insight into the game difficulty and other aspects that enable them to put their personal and comparative findings into the context of this particular game. It is useful mainly in situations when a user plays multiple games. In such a case, the user might want to know why he or she was more successful in one game than in the other, for instance. After group-based training sessions, users are often wondering who was ``the best player'' in the group so that they can further explore his or her tactics and behavior. However, ``the best'' is not easy to define, as seen in tasks \textbf{T11} and \textbf{T12} that deal with various views on ``to be the best''.

Requirements \textsl{R1--R3} delimit the design of interactive visualizations and provide an initial classification for possible interactions. Additional constraints ensue from the available game data. Altogether, they have been considered during the design process (our assumptions on datasets are described in detail in Section~\ref{sec:dataset}). However, there is still a variety of options for a suitable solution. 

To specify user requirements more precisely, we refined \textsl{R1--R3} into particular interactive tasks that are summarized in Table~\ref{tab:tasks}. We aimed to cover various aspects of the high-level requirements. To reach this goal, we built on the analysis of the data available from previous training sessions and the discussion with domain experts who iteratively commented on proposed tasks and voted for them. The resulting list of tasks was, therefore, reached as the consensus of the four aforementioned domain experts. Real meaningfulness of the tasks for players was then verified along with the evaluation of the designed visualizations, as discussed in Section~\ref{sec:method}. We are aware that there can be many other possible tasks associated with the requirements. However, as the goal of this study was to design and validate initial tool serving as an automatically generated visual feedback, we consider the tasks representative.

\begin{table*}[!ht]
\centering
\def\arraystretch{1.1}
\caption{Interactive tasks covering requirements \textsl{R1--R3}.}
\begin{tabular}{crl}
\toprule 
\multirow{6}{*}{\textsl{R1}}
    & T1: & Find out when you finished the game. \\
    & T2: & Find out in which level(s) you reached the lowest score. \\
    & T3: & Find out your final score. \\
    & T4: & Find out how much time you spent in the 2\textsuperscript{nd} level. \\ 
    & T5: & Find out when you advanced from 2\textsuperscript{nd}to 3\textsuperscript{rd} level. \\
    & T6: & Find out in which level you lost most points in the game. \\ \hline
\multirow{3}{*}{\textsl{R2}}
    & T7: & Characterize your score compared to other players. \\
    & T8: & Characterize your time spent by playing compared to other players. \\
    & T9: & Find out the player who reached the closest score to your score. \\ \hline
\multirow{3}{*}{\textsl{R3}}
    & T10: & Find out how much time was assigned for playing the game. \\
    & T11: & Is there somebody who reached a high score in significantly short time? If so insert his Player ID. \\ 
    & T12: & Find out who reached the best score. \\ 
\bottomrule
\end{tabular}
\label{tab:tasks}
\end{table*}

In addition to supporting user requirements \textsl{R1--R3}, one more critical feature had to be considered: The visual feedback has to be intuitive and easy to use since the reflection phase following the training session is often very short (several minutes only), and providing complex visual analytics systems would be counter-productive. The balance between providing a complex set of information, relevant to the trainee and the feedback simplification pose a challenging task.

\section{Visual Design}
\label{sec:visual-design}

We designed two complementary visualizations to provide visual feedback to the trainees of multilevel games. They combine a known clustering and time-based visual approaches to show the score and its development over time in the way that is easy to decode. At present, the visualizations are independent, and each serves for a bit different purpose, presents the data from a different perspective, and on a different level of detail. They were designed to cover together requirements \textsl{R1--R3}.

\subsection{\textsc{Clustering} Visualization}

The first view on the recorded data is presented in the form of a bar chart visualization combined with a scatter plot, as shown in Figure~\ref{fig:teaser} (left).  
It exploits a clustering principle to demonstrate achieved score results and bar chart principles to show time requirements. When designing this visualization, we emphasized the simplicity so that trainees can quickly focus on retrieved score and get a fast overview of their results (\textsl{R1}), results of other trainees (\textsl{R2}), and the overall distribution of results in the game (\textsl{R3}).

The visualization is split into two parts. The upper part with a gray bar includes the overall results of the game, while underneath there are results from individual levels. 

The length of each bar expresses the maximum time for the given level (i.e., the time of the slowest trainee). Bars and timeline on the x-axis are scaled automatically according to the recorded timestamps so that the chart fills the canvas regardless of the game duration. The brighter shade denotes an average time of the level or the game, respectively. Trainees whose results appear in the darker part were faster than the average of all trainees and vice versa. The height of the bars is fixed, although the scoring span can differ in each level. Instead, the scoring span is indicated by numbers next to the y-axis. The fixed height enables users to attract their attention to relevant results in each level or the whole game.

\begin{figure}[!hbp]
  \centering
  \includegraphics[width=\columnwidth]{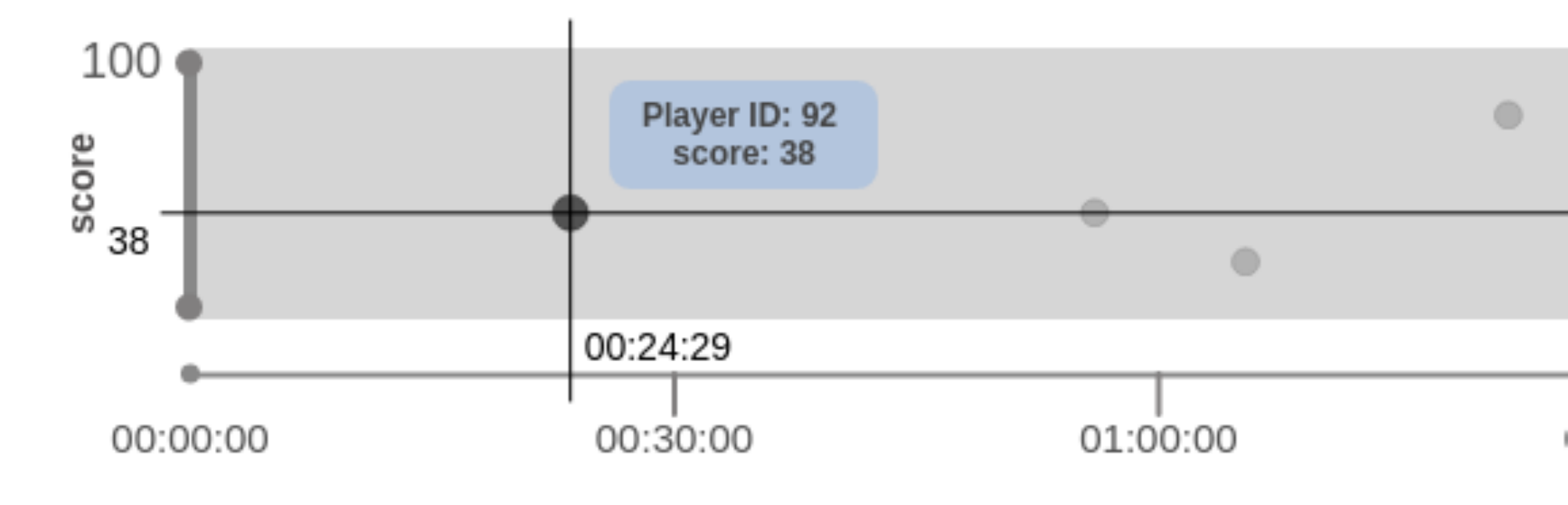}
  \includegraphics[width=\columnwidth]{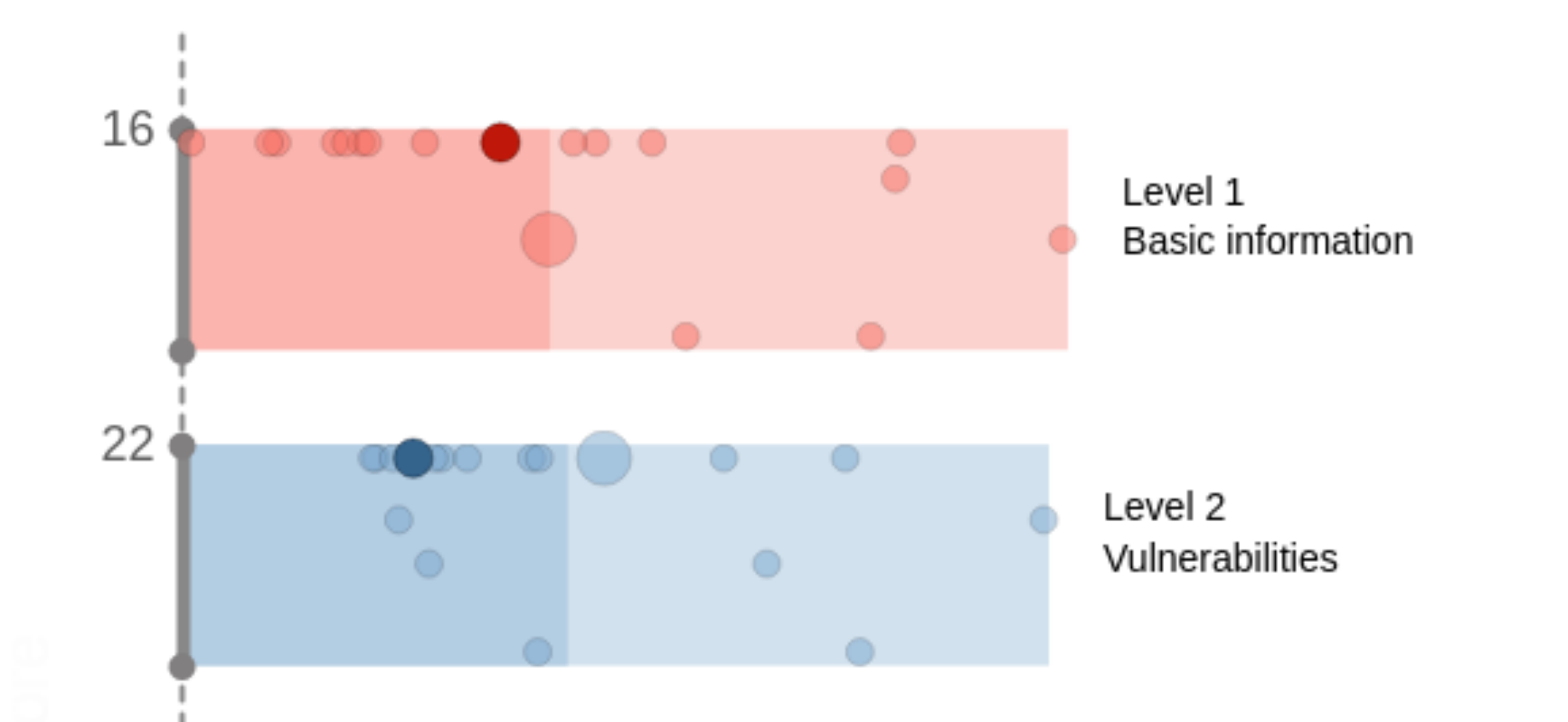}
  \caption{\textsc{Clustering} detail with a selected point on the game overview bar and highlighted corresponding results in level bars.} 
  \label{fig:clustering-detail}
\end{figure}

Results of individual trainees are displayed as small dots. Their vertical and horizontal position in the bars corresponds to their score and time. To support the person-centered view, one of the dots that represents the current trainee is always bulkier than the others. When the mouse pointer hovers over the dot, the corresponding dots of the trainee are highlighted too. This helps to keep track on the individual scores and times of the inspected trainee across multiple levels. Simultaneously, the exact time and achieved score are displayed, as portrayed in Figure~\ref{fig:clustering-detail}. The visualization implements a snapping functionality for attracting the mouse pointer towards the dots to make their selection more comfortable.  

Clusters of points can be used to identify the correlation between time and score visually. Horizontal clusters of points, for instance, reveal users who obtained similar score while vertical clusters show users who spent a similar time in the game levels.

\subsection{\textsc{Timeline} Visualization}

The second visualization, shown in Figure~\ref{fig:teaser} (right),
demonstrates the progress of trainees throughout the game. It is more time-oriented than the previous \textsc{clustering} visualization and also contains more details from the gameplay, including details about hints and penalties that can be shown on demand. 

On the x-axis, there is a timeline, while the y-axis captures score values. The horizontal dashed lines indicate a maximal number of points reachable in the game. In Figure~\ref{fig:teaser}, there is at most 16 points after the first level, 38 points after the second level, 64 points after the third level, and 100 points overall. In contrast to the \textsc{clustering} visualization where the bars are based on recorded timestamps from the game, the striped background of the graph outlines an estimated time for each level of the game. 

\begin{figure}[!hb]
  \centering
  \includegraphics[width=\columnwidth]{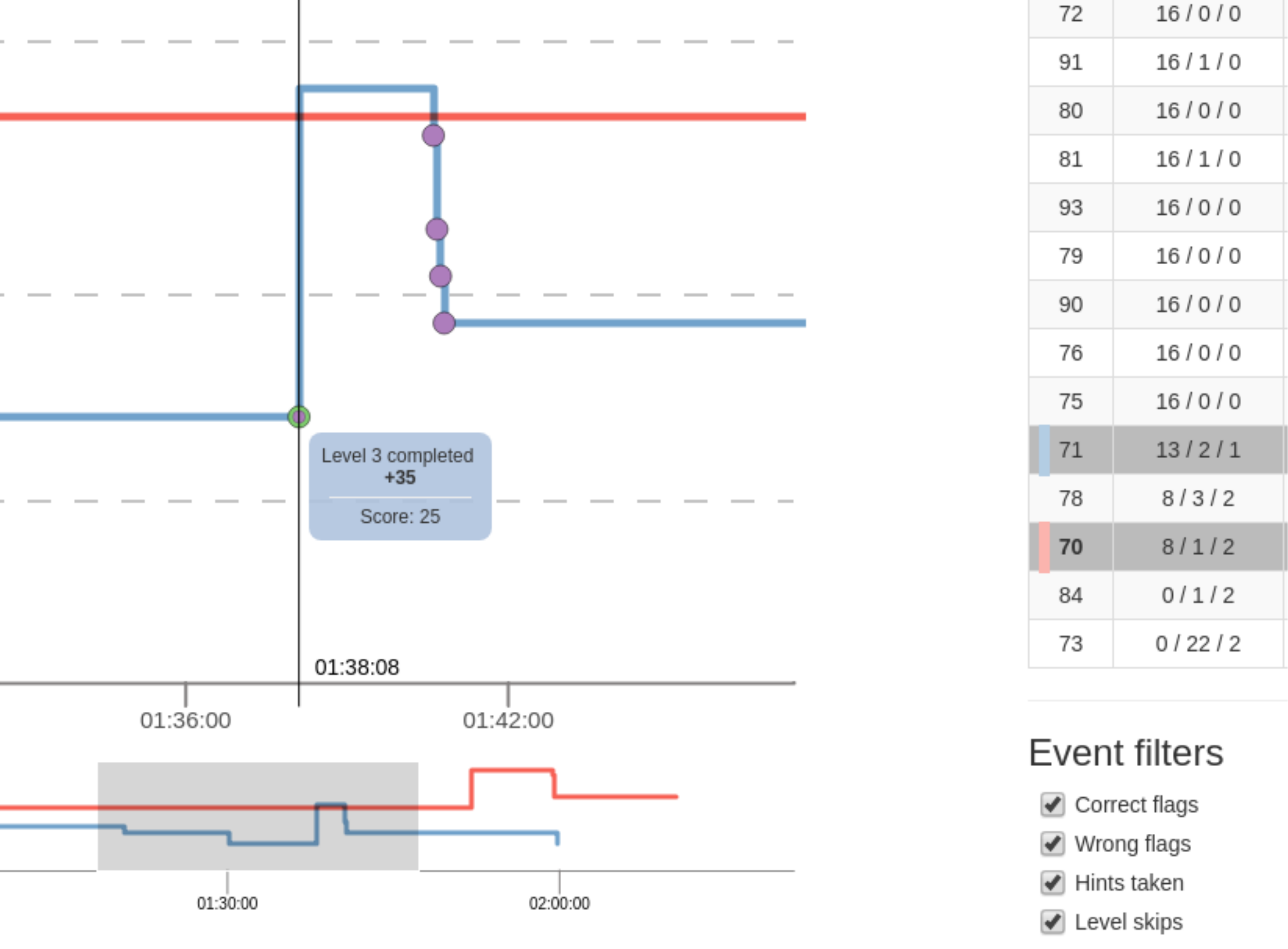}
  \caption{\textsc{Timeline} detail: Zoomed score lines of two trainees with pop-up tooltip of selected event.} 
  \label{fig:timeline-detail}
\end{figure}

Polylines in the visualization, further referred to as \emph{scorelines}, represent fundamental graphical elements showing the development of achieved score of individual trainees. Upon entering a new level, the score increases with the maximum point value for the level and then the \emph{scoreline} significantly ``jumps up'' at this moment. Upon gaining a penalty for providing a wrong flag, taking hints, or skipping the level, the \emph{scoreline} decreases proportionally. Marks of specific events can enrich the scoreline. A pop-up tooltip with event details raises when the mouse cursor hovers the mark, as shown in Figure~\ref{fig:timeline-detail}. Types of events to be shown are controlled by the selection filter next to the main view.

Events in the \emph{scoreline} can be dense. Therefore, we integrated a zooming function into the chart. After zooming, the chart under the main graph provides an overview of zoomed time span and enables the user to adjust a cutout and shift the time span easily.

To support person-centric tasks, the \emph{scoreline} of the current trainee is emphasized. Score lines of individual trainees can be turned on and off by clicking in the adherent table. They have assigned different colors to distinguish score lines of different trainees. The color mapping is shown in the table (color stripes in rows 70 and 71 in Figure~\ref{fig:timeline-detail}, for instance) so that the trainee keeps track of the relationship between table rows and score lines. 

The interactive table also helps in a detailed exploration of results. It provides information about exact scores in individual levels as well as the total score. Users can also quickly compare results by sorting rows according to the selected column (level or final score). 

\subsection{Implementation}

Both visualizations are implemented as Angular modules using the D3.js library for drawing. A demo version with test data is available online at \url{http://kypo-summer.surge.sh/}. 
The visualizations are adapted for independent testing and evaluation. Although they contain a dummy dataset (which is different from the data used for this usability study), it is also possible to upload new game definitions (number of game levels, their names, etc.) and corresponding logs capturing the gameplay data of trainees. The game definition and event logs used in this usability study are available in supplementary materials \protect\footnotemark[2].

We are currently integrating the visualizations into our KYPO Cyber Range~\cite{KYPO} -- a cyber exercise and research environment that is used for the organization of hands-on cybersecurity training~\cite{ITICSE18kypolab}. When the integration is done, the data will be available at runtime, and the visualizations will become a part of the learning process.

\section{Usability Study} 
\label{sec:method}

In this section, we describe the details of the usability study we held to evaluate the usability of the visual feedback. We decided to conduct a within-subject user evaluation with the attendees of Summer School in Cyber Security 2018 to mimic a real-world use case of the visualizations. The user study was included as a part of the program. Participants were finalists of high-school Cyber Security Competition 2018.

\subsection{Hypotheses}

We have formulated three hypotheses for the evaluation. They address the meaningfulness of our user requirements and related tasks, the usefulness of the visualizations in solving the tasks, and the identification of strengths and weaknesses of our approach. They are defined as follows:

\textbf{\textsf{H1}: Requirements \textsl{R1--R3} are meaningful and useful for trainees.} User requirements and their corresponding tasks were distilled from the discussion with domain experts -- game designers and organizers of CTF games. The goal of this hypothesis is to verify that the requirements are also meaningful and useful for trainees and that they sufficiently cover their interests. To verify this hypothesis, the players rated the meaningfulness of individual tasks. 

\textbf{\textsf{H2}: The visual feedback is useful in providing insight into the tasks of \textsl{R1--R3}.} This hypothesis should verify our assumption that the visual feedback provides straightforward and easy to decode way for seeking relevant information. Verification of this hypothesis was based on the qualitative and quantitative evaluation, during which the participants solved tasks and rated their difficulty.

\textbf{\textsf{H3}: Some visualizations or their parts are more useful for specific tasks of \textsl{R1--R3} than others.} The \textsc{Clustering} and \textsc{Timeline} visualizations were designed as complementary, providing different views on the data with different level of detail. However, most tasks \textbf{T1}--\textbf{T12} can be solved by both of them, in an easier or more difficult way. The hypothesis \textsf{H2} should uncover the usefulness of the visual feedback as a whole, regardless of which visualization was used to solve the task. On the contrary, this hypothesis aims to verify whether some views or parts of the visual feedback fit better to some tasks or user requirements than others. Our goal is not only to confirm or reject the hypothesis but to identify such tasks and visualizations. To reveal this type of information, we asked for the usefulness of individual views for solving particular tasks.

\subsection{Participants}
Out of 16 attendees of the summer school, 12 senior high-school students (1 female, 11 male) participated in the study. All of them were between 16 and 19 years old, with normal or corrected-to-normal vision. None of them was color blind. All of them were daily users of smartphones and computers (laptops or desktops). Nine of them considered themselves as experienced users familiar with cybersecurity topics, one as a complete newcomer, one as a novice user, and one as a security professional. Although none of them was a native English speaker, they were proficient enough to understand the English questionnaire.

\subsection{Environment}
The evaluation was conducted on all-in-one computers with FullHD displays ($1920\times1080$ pixels) running Windows 10 and the latest stable Google Chrome as an internet browser. The browser was maximized the whole time. The participants used the same computers as in the previous activities of Summer School. 

We used the LimeSurvey online questionnaire tool for presenting the informed consent form, instructions, task assignments and complementary questions (task meaningfulness, difficulty, and visualization preference for each of the two). The tasks were organized one per screen and accompanied with supplementary questions. 

\subsection{Procedure}

The participants engaged with the CTF game described in Section~\ref{sec:dataset}. They utilized a web interface of the KYPO Cyber Range~\cite{KYPO} to play the game. The feedback visualizations were not available to players during the game. The rough time assigned for the CTF game was 90 minutes. When the time ran out, the participants who did not finish had to terminate the game using the ``Skip level'' function.

Then, there was a 20-minute refreshment break during which the operators prepared the LimeSurvey questionnaire, and set up the feedback visualizations. The latter step included loading real logs from the CTF game. The questionnaire and the two visualizations were opened in separate tabs within the Google Chrome browser. Since \textsc{clustering} and \textsc{timeline} visualizations were designed as complementary, the participants could use both to accomplish the tasks. Therefore, the participants had to switch between tabs when they solved the tasks. The names of the visualizations were also displayed in tab labels to avoid unintentional terminology mismatch between them.

After the break, the operators explained the purpose of the experiment and the user study procedure comprising of three parts. We explicitly asked the participants not to collaborate among each other.

First, the participants were introduced with the two visualizations, and they had up to 10 minutes to familiarize with both of them. Next, the 12 tasks described in Table~\ref{tab:tasks} were assigned to them. To mitigate the ordering bias, we randomized the order of the questions for each participant. Last but not least, the participants answered several demographic questions. In total, we reserved 50 minutes for the user study. One of the operators was present the whole time to provide support with technical issues.

\subsection{Results} 

This section presents the result of a quantitative and qualitative evaluation of the data acquired from the usability study. The questionnaire and collected answers are included in supplementary materials\protect\footnotemark[2].

The independent variables included in the study were tasks (12) and visualizations (2). For these, we measured two dependent variables: task correctness and ordinal data from 6-point Likert scales focused on usability of each visualization for a particular task (1 = Absolutely useless, 6 = Absolutely useful), difficulty of the task and its meaningfulness (both: 1 = Strongly disagree, 6 = Strongly agree), as discussed in what follows. We obtained 192 trials (12 participants $\times$ (12 tasks + task meaningfulness + task difficulty + visualization preference)) from the usability study. 

\subsubsection*{\textsf{H1} -- Meaningfulness of Requirements}

\begin{figure}[!ht]
  \centering
  \includegraphics[width=0.7\linewidth]{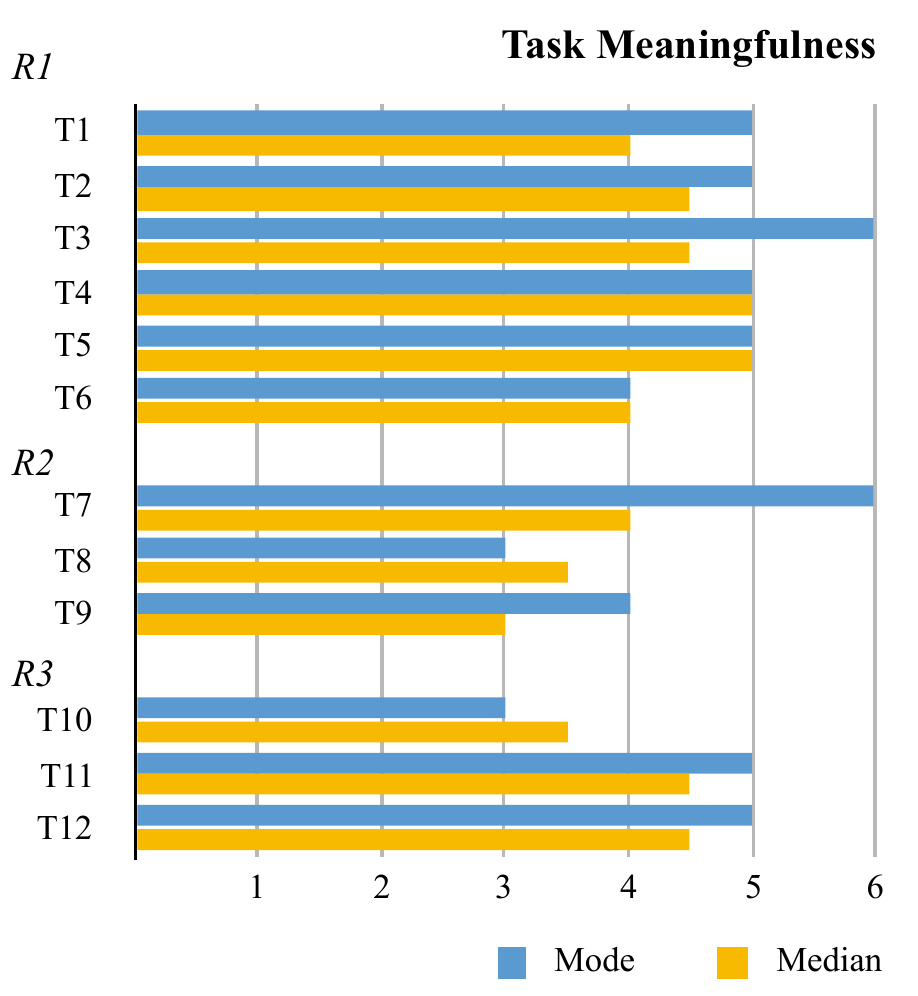}
  \caption{Evaluation of the tasks meaningfulness. \textit{Question: The task was meaningful.} Higher score is better (1 = strongly disagree, 6 = strongly agree).} 
  \label{fig:meaningfulness}
\end{figure}

To verify that the design requirements \textsl{R1}--\textsl{R3} were chosen reasonably and the tasks \textbf{T1}--\textbf{T12} reflect users interests, we analyzed answers to the question \emph{The task was meaningful} that has been asked after each task. Figure~\ref{fig:meaningfulness} presents the median and mode values. The overall score provided by participants is positive, however, not significantly. $Median=4$ (= somewhat agree) for all three requirements. $Mode=5$ for \textsl{R1} and \textsl{R3}, $mode=3$ for \textsl{R2}.



We can notice higher scores in case of \textsl{R1} and \textsl{R3} compared to \textsl{R2}. It means that participants considered tasks in these two categories more meaningful when reflecting their gameplay.
Friedman test reveals no statistical significance ($\alpha=.05$) among the tasks within the same requirement group: \textsl{R1} ($\chi^2=4.97, df=2, p=0.42$), \textsl{R2} ($\chi^2=4.96, df=2, p=0.084$), and \textsl{R3} ($\chi^2=1.51, df=2, p=0.459)$. This observation confirms hypothesis \textsf{H1} and our assumption that the tasks reflect users interests.

\subsubsection*{\textsf{H2} -- Usefulness of Visualizations}

Both \textsc{clustering} and \textsc{timeline} visualizations were designed as complementary. Therefore, our goal was not to compare usefulness head-to-head but evaluate their usefulness for completing the tasks. 

First, we analyzed every task individually to determine (in)correct responses and their ratio. Since the data was produced by participants while playing the game, thus are not synthetic, some of the tasks do not have a simple one-value correct solution. Moreover, in time-related questions, there can be inaccuracy in answers caused by many reasons, e.g., approximate mouse location on the visualization, ignoring seconds by the user, etc. Due to these facts, we checked individual responses and categorized them into three groups: wrong, correct, and partially correct. This assessment was reached as the consensus of the authors of the paper. 

\begin{figure}[!ht]
  \centering
  \includegraphics[width=\linewidth]{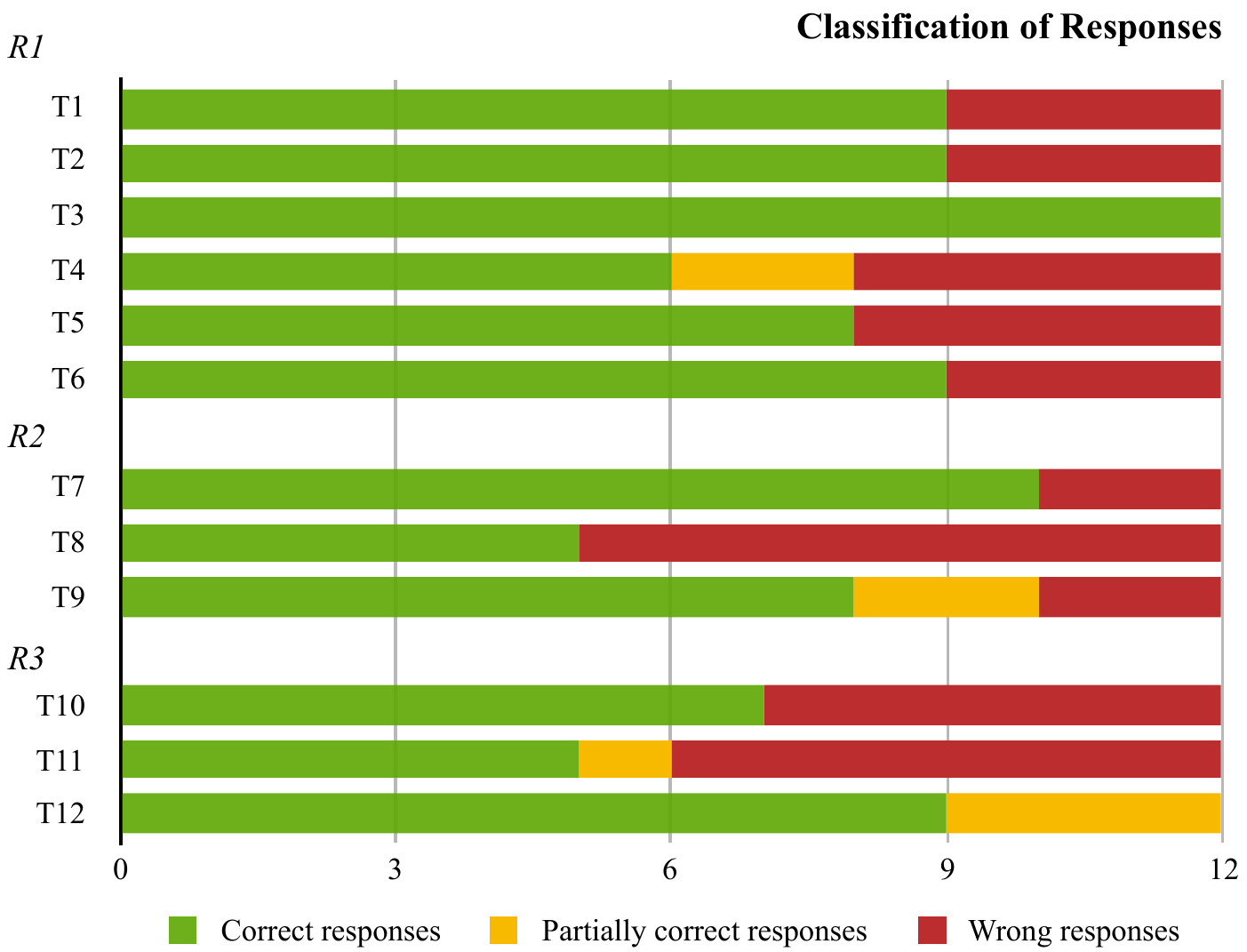}
  \caption{Responses classification for each task.} 
  \label{fig:responses}
\end{figure}

Figure~\ref{fig:responses} presents resulting (classified) responses. The overall combined success rate (including both correct and partially correct responses) is $73\,\%$ with eight correct responses per task on average. Therefore, we can conclude that the trainees were successful in performing tasks in general.

We identified four occurrences of partially correct responses. The imprecise answers when the reported time was rounded to the nearest minute (e.g., 0:15:00 instead of 0:14:59) in \textbf{T4}. In \textbf{T9} and \textbf{T12}, there are incomplete answers when participants reported only a subset of multiple correct answers (e.g., two participants reached the same best score but only one of them was reported). In \textbf{T11}, it was due to unclear data when there was no strong evidence with multiple correct options.

\begin{figure}[!ht]
  \centering
  \includegraphics[width=0.75\linewidth]{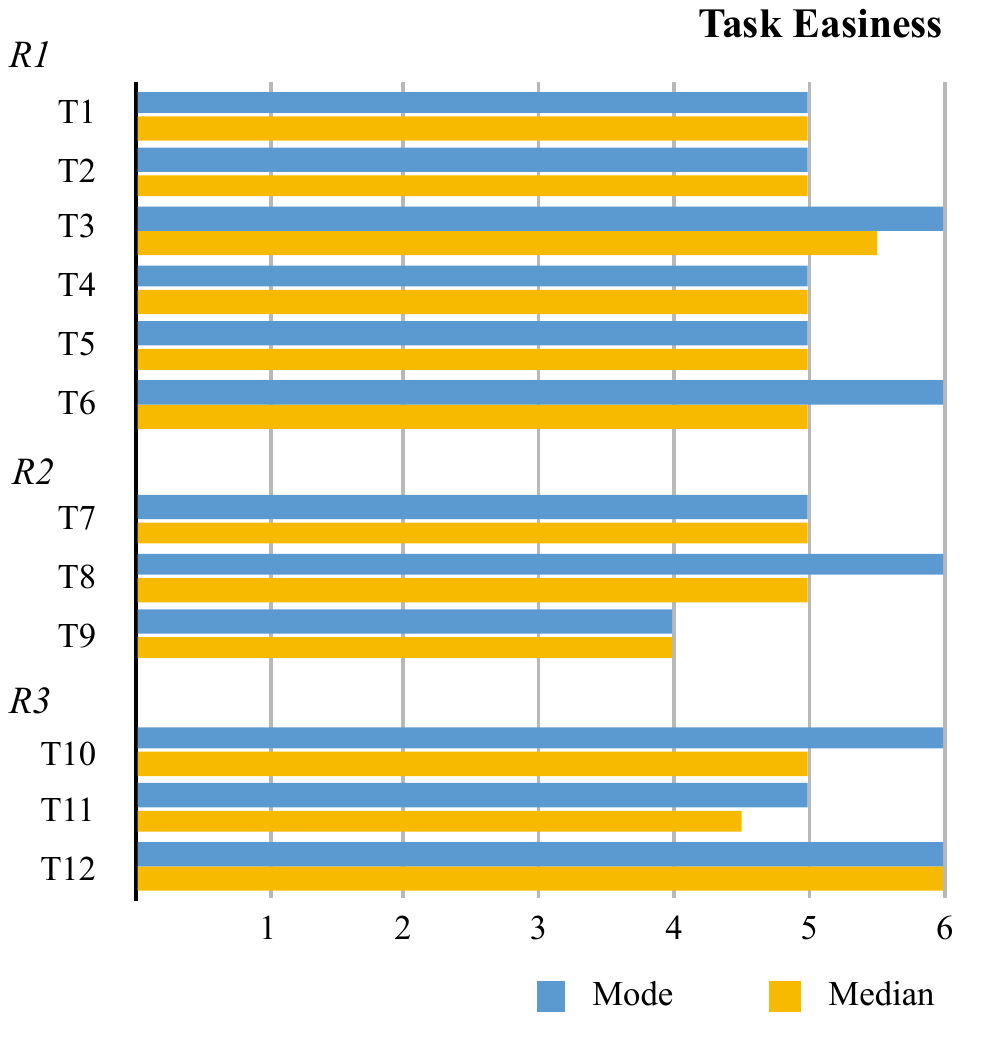}
  \caption{Evaluation of the tasks easiness. \textit{Question: The task was easy to complete (using the visualizations).} Higher score is better, (1 = Strongly disagree, 6 = Strongly agree).} 
  \label{fig:easiness}
\end{figure}

Figure~\ref{fig:easiness} depicts the results of the qualitative evaluation of the Hypothesis \textsf{H2} presented by the question \textit{The task was easy to complete (using the visualization)}. Regardless of the actual number of wrong responses, the overall feedback was that the tasks were easy to solve with the visualizations (\textit{mode = 5, median = 5}). 
Friedman test does not reveal any statistical significance among the tasks of \textsl{R1} ($p=0.27$). However, the two remaining groups have significant difference among their tasks. \textsl{R2} ($\chi^2=10.207, df=2, p<.05$), \textsl{R3} ($\chi^2=6.5, df=2, p<.05$). A post-hoc analysis using Conover's F test reveals that \textbf{T9} (\textit{Find out the player who reached the closest score to your score.}) is statistically significantly difficult opposed to \textbf{T7} (\textit{Characterize your score compared to other players.}) and \textbf{T8} (\textit{Characterize your time spent by playing compared to other players.}). Likewise, there was a statistically significant difference in difficulty between \textbf{T11} (\textit{Find out in which level you lost most points in the game.}) and \textbf{T12} (\textit{Find out who reached the best score.}) in \textsl{R3}. Despite these observations, both the median and the mode of these two tasks is still greater or equal to 4 (= slightly agree).
Therefore, we conclude that the hypothesis \textsf{H2} was confirmed. The use of visualizations supports trainees' understanding and orientation in the game data. 

\subsubsection*{\textsf{H3} -- Preferences in Using Visualizations}

\begin{figure*}[ht!]
  \centering
  \includegraphics[width=.9\linewidth]{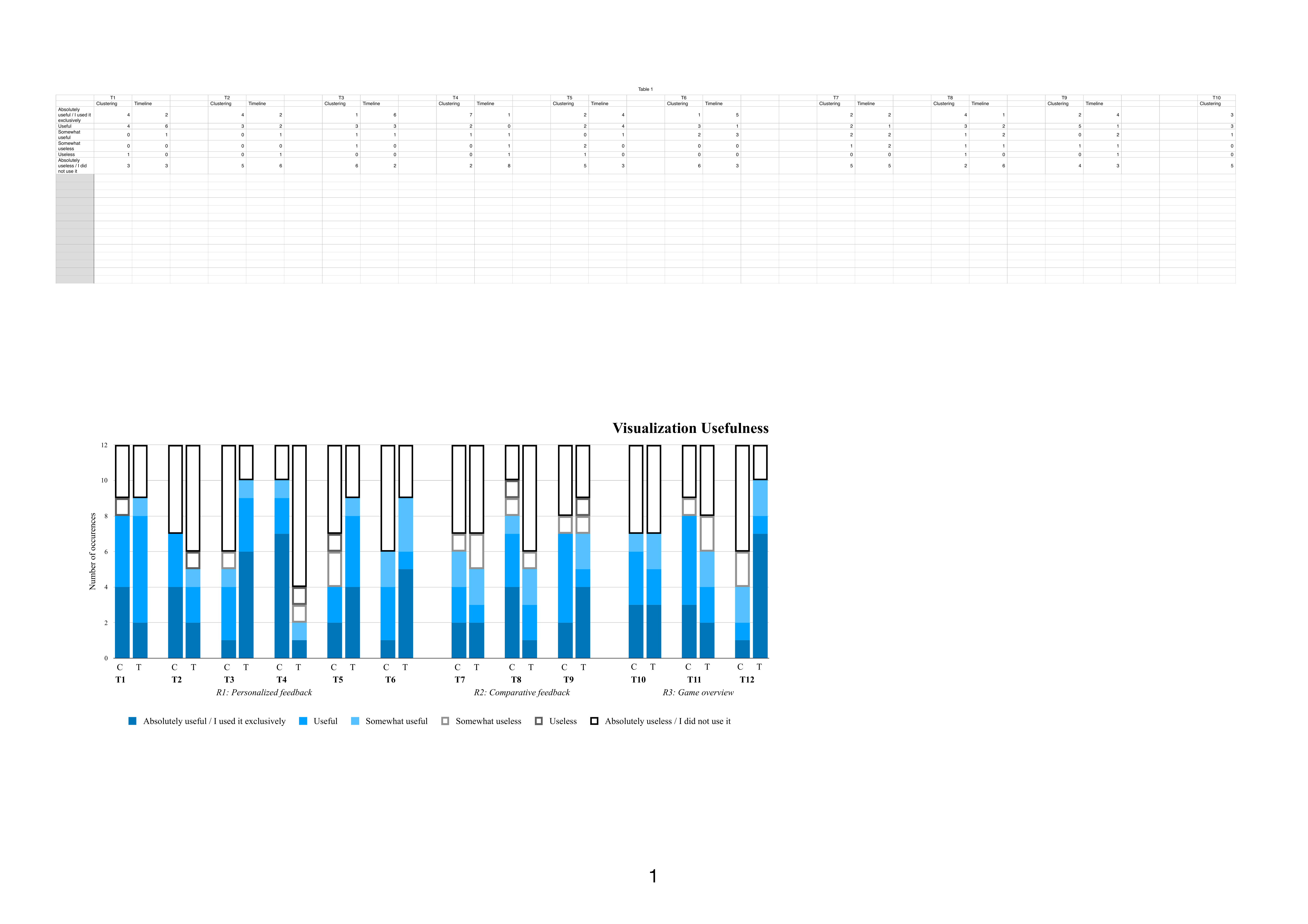}
  \caption{Evaluation of the visualization usefulness with respect to the tasks. \textit{Question: Evaluate the usefulness of the [\textsc{clustering} | \textsc{timeline}] visualization for the task.}; C = \textsc{Clustering} visualization, T = \textsc{Timeline} visualization.} 
  \label{fig:usefulness}
\end{figure*}

Our last hypothesis focuses on determining preferred visualizations for individual tasks. 
Figure~\ref{fig:usefulness} presents the complete results obtained from a pair of questions focused on usefulness evaluation of visualizations with regards to the tasks. Although we find only slight differences between most of them, it does not apply for a subset of four tasks where we can observe stronger preference either for \textsc{timeline} (\textbf{T3, T5, T12}) or \textsc{clustering} (\textbf{T4}) visualization.
To confirm our assumption on statistical significance, we performed one-tail Wilcoxon Matched Pairs Signed-rank Test. According to our expectations, there is no statistically significant preference for tasks \textbf{T1}, \textbf{T2}, \textbf{T6}--\textbf{T11}. Even \textbf{T5} is not significant $(W=17, p=0.28)$. The three remaining tasks have significant preference for \textsc{timeline} visualization: \textbf{T3} (W = 8), \textbf{T12} (W = 8); and \textsc{clustering} visualization: \textbf{T4} (W = 7.5). 

We confirm hypothesis \textsf{H3} since there are at least four tasks, where we observed a statistically significant preference for either of the visualizations. However, only limited conclusions can be drawn for this hypothesis due to common tied values in our data. As a result, the sample size was often lowered by up to 5 samples (e.\,g., \textbf{T2, T7}). Also, the Wilcoxon test is known to be less sensitive when the sample size is very low ($N<10$) and any difference that is statistically significant will have to be huge. Thus, further inspection with a larger sample is still needed. 

\section{Discussion}
\label{sec:discussion}

In this section, we summarize the results of the usability study, discuss limitations and lessons learned.

\subsection{Summary of Evaluation Results}

\textit{\textbf{Design requirements are correct, and the tasks reflect user interests.}} The evaluation confirmed that the user requirements \textsl{R1}--\textsl{R3} distilled from interviews with four domain experts are meaningful and useful for trainees, and related tasks reflect user interests. Nevertheless, participants considered tasks of \textsl{R1} (personalized feedback) and \textsl{R3} (a brief overview of the overall game results and features) more meaningful when reflecting their gameplay than \textsl{R2} (comparative feedback).

\textit{\textbf{Visualizations support trainees in the understanding of results.}} In general, trainees were able to complete given tasks correctly and, according to their response, tasks were easy to solve with the visual feedback. The evaluation revealed that some tasks were more difficult to solve than others. We aim to better support tasks identified as difficult in further research.

\textit{\textbf{We did not find that any of the visualizations would better support any of the user requirements.}} On the other hand, we identified specific tasks for which one of the visualizations might be more appropriate. However, the results are uncertain due to data limitations. Further inspection with a larger sample is still needed to validate them.

\textit{\textbf{We have received several suggestions for improvement.}} Trainees are not expected to interact with the feedback tool often. Therefore, it is important to reflect user experience in the design of the visualization tools so that they get familiar with them easily. We received particular suggestions and bug reports from the evaluation that we will reflect in the future development of the visual feedback. One participant noted that ``the marking signifies the last change in score, but it's NOT the time I stopped playing''. Three participants reported inconsistency in data presentation between the visualizations when the displayed time could vary by up to one second due to rounding off raw time-stamps. One participant suggested an improvement for the \textsc{timeline} visualization: ``There should be some buttons to select or deselect all players at once. Now the user has to click on each player individually.''

\subsection{Usability Study Limitations}

\textit{\textbf{Low number of participants.}} We decided for the field usability study since we wanted to reach as realistic settings as possible, which would be only hardly achievable in a controlled experiment. According to Rubin et al.~\cite{Rubin1994}, a truly experimental usability test achieving statistically valid results should be conducted with a minimum of 10 to 12 participants per condition. And although the sample size of 12 participants is not unusual in similar studies in general, we are aware of this weakness for claiming a strong confirmation of our findings. Since research has shown that a sample size of 4 to 5 participants can expose about 80\,\% of usability issues~\cite{Rubin1994}, and since the hypotheses were confirmed, and initial outcomes are positive, we consider the results as promising and entitling us to elaborate the feedback visualizations further.

\textit{\textbf{Dataset limitations.}} The dataset used in the evaluation was not synthetic. Thus some of the tasks defined prior to the experiment do not have clear ``one-value only'' answers. There were multiple correct solutions or data was unclear. As a result, some of the tasks were more difficult to solve, or the solution was not straightforward (e.g., \textbf{T9}, \textbf{T11}). We also faced three technical issues on the cyber range infrastructure when some of the game events were not recorded. As a result, the visualizations did not reflect the real-world situation of three participants, which was confusing for them and probably affected some of the responses in \textbf{T8}.

\textit{\textbf{Ambiguous responses to visualization preference.}} We identified eight ambiguous responses where, despite the instructions, participants tick \textit{Absolutely useful / I used it exclusively} in one visualization and different option than \textit{Absolutely useless / I did not use it} in the other. Three of the responses were from the same participant. Some of the participants also noted that they use the table (which is, in fact, a part of the \textsc{timeline}) instead of visualization itself. Since we plan to repeat the usability study, we are going to revise the way of presenting the questions to reduce the ambiguity and extract more qualitative information on the actual use of visualizations.

\subsection{Observations and Lessons Learned}

\textit{\textbf{Trainees prefer exploration of personal results to the overall game results and comparison with others.}} Detailed inspection of the results revealed that the participants were primarily interested in their score (tasks of \textsl{R1}), followed by the overall awareness of the game (tasks of \textsl{R3}). We assume that the primary objective of a player is to get insight into his/her gameplay (a score development, the time they spent playing the game and its parts). Further, they are interested in the overall game situation without bothering too much with a detailed comparison with others (tasks of \textsl{R2}). We assume that the comparative perspective is meaningful only in specific cases -- e.g. when two friends want to compare their scores. Our initial findings and confirmation of \textsf{H1} open a new research topic of determining the essential information for trainees and their additional support in post-game feedback visualizations. Since this is far beyond the scope of this paper, we leave it for our follow-up work.

\textit{\textbf{Easy to decode design is not mandatory.}} Our preliminary expectations that a much simpler \textsc{clustering} visualization would be more useful and used than more complex \textsc{timeline} visualization have not been confirmed. Even though both visualizations were designed for a different type of tasks, there is considerable overlap in their capabilities in answering the same questions. For example, both of them offer a straightforward way of finding the final score, and the scores reached in individual levels. However, the study did not reveal preference of the \textsc{clustering} to \textsc{timeline}. Data of only four of the tasks reported the preference for either of them. Although the intuitiveness of the design is still mandatory due to the restrictions of the reflection phase, the easy to decode design seems not to be essential, as we expected. The users are willing to use a more complex variant if they provide more details. There were also a participant who prefers simple presentation in tabular form from fancy visualizations.

\textit{\textbf{Trainees tend to perceive time subjectively.}} The task \textbf{T8} (\textit{Characterize your time spent by playing compared to other players.}) was ranked with five options from \textit{I was one of the slowest players} to \textit{I was one of the fastest players}. Answers to this question included the most number of wrong responses (58\%). Participants either under- or overestimated their finish time compared to others. Unfortunately, none of them explicitly commented on why. We assume that participants reflected their impression from real-world time (someones started playing a little earlier, some later) rather than precise relative game time represented in visualizations.

\textit{\textbf{Do not mix time spans with a different meaning,}} even if they are in separate visualizations and explained by a legend. We noted yet another confusion with the time-related tasks. Whereas \textsc{clustering} visualization shows average and total time based on the trainees' data, \textsc{timeline} visualization displays estimated time for each level (colored diagonal stripes on the background in Figure~\ref{fig:teaser}). \textbf{T10} (\textit{Find out how much time was assigned for playing the game.}) targets the latter one. All the participants who made a mistake indicated the maximal time from the \textsc{clustering} visualization instead. 

\textit{\textbf{Users can interpret game results subjectively.}} Game results consist of a score and time in which the score was reached. In general, the best players are considered those who achieved the best score. Nevertheless, also trainees who did not get the absolute best score but solved the task quickly can be considered as very successful. 
The task \textbf{T11} (\textit{Is there somebody who reached a high score in significantly short time? If so, insert his Player ID.}) was to reveal such trainees. However, the task has a considerably high number of wrong responses (50\%). From the detailed inspection of the results, we found out that the biggest issue is unclear data that makes the task difficult. There was no single recognizable answer in the data. We (the authors) agreed that the correct response is the player who finished as the first one. He was also one of the two players with the highest score. Most of the participants (in five cases) marked the same player. One of the participants marked both players with the highest score (i.e., partially correct response). The rest of the participants marked different ones, usually those who finished earlier but had a rather low score. A solution to this problem could be to include data storytelling principles so that users immediately see what is important by providing a ``narrative'', as discussed in~\cite{echeverria2018}, for instance.

\begin{figure}[!ht]
  \centering
  \includegraphics[width=\linewidth]{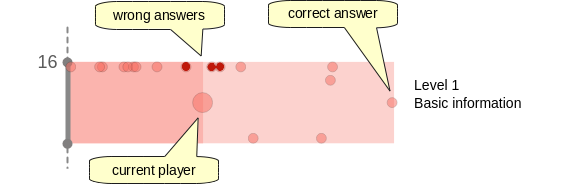}
  \caption{Clustering weakness -- the users tend to select closer neighbors while omitting the distant ones.} 
  \label{fig:clustering-weakness}
\end{figure}

\textit{\textbf{The clustering approach can be confusing in specific cases}} Responses of \textbf{T9} (\textit{Find out the player who reached the closest score to your score.}) revealed one weakness in the design of the \textsc{clustering} visualization. One participant, who used this visualization solely and provided the wrong player ID, inspected only the closest neighborhood where he found the wrong answer. The correct one was placed too far right from his position since this participant was one of the slowest, as illustrated in Figure~\ref{fig:clustering-weakness}. In this situation users tend to prefer closer neighbours to more distant points. A solution to this problem could be to gradually highlight those who are in, e.g., 10-25-50\,\% dispersion around both horizontal and vertical dimensions. Additionally, the pop-up tooltip (raised when the mouse cursor hovers over the dot) could be extended with Top 3 better and Top 3 worse trainees regarding the observed one.

\section{Conclusion and Future Work}
\label{sec:conclusion}

The work we presented in this paper focuses on improving post-game feedback for players (trainees) of serious multi-level cybersecurity games by using interactive visualizations. The feedback is one of the critical phases of the learning process. However, the way of presenting the results is often limited to plain scoreboards presenting only the total scores and generalized feedback on the most common issues observed during the gameplay. We improve the user experience through design and implementation of two interactive visualizations. The visualizations improve overall situational awareness and insight into the gameplay. Also, they provide a straightforward way for comparison of individual trainees. A demo version with test data is available online at \url{http://kypo-summer.surge.sh/}. 

We collaborated with four cybersecurity education experts on defining visualization requirements. Together, we formulated a set of tasks that address three areas of trainees' interest: gameplay overview, person-centered feedback, and comparative feedback. Through the usability study held with participants of Summer School in Cybersecurity, we evaluated three hypotheses related to (a) meaningfulness of the user requirements and related tasks, (b) usefulness, and (c) preference of the visualizations regarding the tasks. All three hypotheses were confirmed, but due to a low sample size (only 12 participants), some of the conclusions will be addressed in our future work. Usability study also confirmed the practical usefulness of the visualizations and pointed out several topics worth further investigation. Namely, defining the set of preferred information for trainees (such as correct or incorrect attempts) and their additional support in feedback visualizations. Further, we collected valuable feedback on the strengths and weaknesses of the visualizations, which we want to address in the implementation. Although the results are inconclusive, the usability study also revealed a preference for one of the visualizations concerning the specific tasks. 

Our work is still in progress. In the future, we intend to further improve the previously mentioned following aspects. Our overall goal is to provide, through interactive visualizations, personalized feedback to increase gained knowledge. 
Our ongoing work on integrating both visualizations into the cyber range web portal is just the first step. The integration also implies interconnecting and extending the interaction capabilities of both visualizations. For example, selecting a trainee in one visualization will highlight the same relevant data in the other. We also want to integrate the visualizations into the gameplay and extend their capabilities toward personalized post-level feedback. The aim is to provide an instant user-specific overview of strengths and weaknesses based on the user's actions.
Our research will include experimental evaluation to confirm and strengthen the initial conclusions from the presented usability study as well as evaluation of new features. Hands-on cybersecurity training is a part of regular university lectures.  Therefore, we prefer running more extensive field study over standard laboratory evaluation on synthetic data. 

The visualizations were designed to provide timely feedback to trainees. However, there are other users involved in the CTF life cycle that would benefit from an interactive exploration of CTF results. The visualizations discussed in this paper enable \emph{supervisors} (the educators who oversee the game) to reflect the overall results so that they can assess their interventions during the game sessions. Similarly, \emph{game designers} (the authors of the content) can utilize the visualizations in their workflow to evaluate game parameters. We will address these issues in future research as well.

\acknowledgments{
The authors would like to thank the cybersecurity experts for providing the feedback and invaluable insight into the target domain. This research was supported by ERDF ``CyberSecurity, CyberCrime and Critical Information Infrastructures Center of Excellence'' (No. CZ.02.1.01/0.0/0.0/16\_019/0000822). Computational resources were provided by the European Regional Development Fund Project CERIT Scientific Cloud (No. CZ.02.1.01/0.0/0.0/16\_013/0001802).}

\bibliographystyle{abbrv-doi}
\bibliography{references}

\begin{thebibliography}{10}

\bibitem{abt1987}
C.~C. Abt.
\newblock {\em Serious Games}.
\newblock University Press of America, 1987.

\bibitem{arnab15}
S.~Arnab, T.~Lim, M.~B. Carvalho, F.~Bellotti, S.~De~Freitas, S.~Louchart,
  N.~Suttie, R.~Berta, and A.~De~Gloria.
\newblock {Mapping Learning and Game Mechanics for Serious Games Analysis}.
\newblock {\em British Journal of Educational Technology}, 46(2):391--411,
  2015.

\bibitem{bellotti13}
F.~Bellotti, B.~Kapralos, K.~Lee, P.~Moreno-Ger, and R.~Berta.
\newblock {Assessment in and of Serious Games: An Overview}.
\newblock {\em Adv. in Hum.-Comp. Int.}, 2013:1:1--1:1, Jan. 2013. doi: {{%
10\hspace{.1pt}\discretionary{.}{%
}{.}\hspace{.4pt}1155\discretionary{/}{%
}{/}2013\discretionary{/}{%
}{/}136864}}


\bibitem{bodily2017}
R.~Bodily and K.~Verbert.
\newblock Trends and issues in student-facing learning analytics reporting
  systems research.
\newblock In {\em Proceedings of the seventh international learning analytics
  \& knowledge conference}, pp. 309--318. ACM, 2017.

\bibitem{boopathi2015}
K.~Boopathi, S.~Sreejith, and A.~Bithin.
\newblock {Learning Cyber Security Through Gamification}.
\newblock {\em Indian Journal of Science and Technology}, 8(7):642--649, 2015.

\bibitem{cao2015}
N.~Cao, L.~Lu, Y.-R. Lin, F.~Wang, and Z.~Wen.
\newblock Socialhelix: visual analysis of sentiment divergence in social media.
\newblock {\em Journal of Visualization}, 18(2):221--235, 2015.

\bibitem{Cao2016}
N.~{Cao}, C.~{Shi}, S.~{Lin}, J.~{Lu}, Y.~{Lin}, and C.~{Lin}.
\newblock Targetvue: Visual analysis of anomalous user behaviors in online
  communication systems.
\newblock {\em IEEE Transactions on Visualization and Computer Graphics},
  22(1):280--289, Jan 2016. doi: {{%
10\hspace{.1pt}\discretionary{.}{%
}{.}\hspace{.4pt}1109\discretionary{/}{%
}{/}TVCG\hspace{.1pt}\discretionary{.}{%
}{.}\hspace{.4pt}2015\hspace{.1pt}\discretionary{.}{%
}{.}\hspace{.4pt}2467196}}


\bibitem{picoctf2}
P.~Chapman, J.~Burket, and D.~Brumley.
\newblock {PicoCTF: A Game-Based Computer Security Competition for High School
  Students}.
\newblock In {\em 2014 {USENIX} Summit on Gaming, Games, and Gamification in
  Security Education (3GSE 14)}. {USENIX} Association, San Diego, CA, 2014.

\bibitem{chung2017}
K.~Chung.
\newblock Live lesson: Lowering the barriers to capture the flag administration
  and participation.
\newblock In {\em 2017 {USENIX} Workshop on Advances in Security Education
  ({ASE} 17)}. {USENIX} Association, Vancouver, BC, 2017.

\bibitem{Corrin2015}
L.~Corrin and P.~de~Barba.
\newblock How do students interpret feedback delivered via dashboards?
\newblock In {\em Proceedings of the Fifth International Conference on Learning
  Analytics And Knowledge}, LAK '15, pp. 430--431. ACM, New York, NY, USA,
  2015. doi: {{%
10\hspace{.1pt}\discretionary{.}{%
}{.}\hspace{.4pt}1145\discretionary{/}{%
}{/}2723576\hspace{.1pt}\discretionary{.}{%
}{.}\hspace{.4pt}2723662}}


\bibitem{ctftime.org}
{CTFtime team}.
\newblock {CTFtime.org / All about CTF (Capture The Flag)}.
\newblock \url{https://ctftime.org/}.
\newblock (Accessed on 2018-06-13).

\bibitem{deFreitas2017}
S.~de~Freitas, D.~Gibson, V.~Alvarez, L.~Irving, K.~Star, S.~Charleer, and
  K.~Verbert.
\newblock How to use gamified dashboards and learning analytics for providing
  immediate student feedback and performance tracking in higher education.
\newblock In {\em Proceedings of the 26th International Conference on World
  Wide Web Companion}, WWW '17 Companion, pp. 429--434. International World
  Wide Web Conferences Steering Committee, Republic and Canton of Geneva,
  Switzerland, 2017. doi: {{%
10\hspace{.1pt}\discretionary{.}{%
}{.}\hspace{.4pt}1145\discretionary{/}{%
}{/}3041021\hspace{.1pt}\discretionary{.}{%
}{.}\hspace{.4pt}3054175}}


\bibitem{doupe}
A.~Doup{\'e}, M.~Egele, B.~Caillat, G.~Stringhini, G.~Yakin, A.~Zand,
  L.~Cavedon, and G.~Vigna.
\newblock {Hit 'Em Where It Hurts: A Live Security Exercise on Cyber
  Situational Awareness}.
\newblock In {\em Proceedings of the 27th Annual Computer Security Applications
  Conference}, ACSAC '11, pp. 51--61. ACM, New York, NY, USA, 2011. doi: {{%
10\hspace{.1pt}\discretionary{.}{%
}{.}\hspace{.4pt}1145\discretionary{/}{%
}{/}2076732\hspace{.1pt}\discretionary{.}{%
}{.}\hspace{.4pt}2076740}}


\bibitem{echeverria2018}
V.~Echeverria, R.~Martinez-Maldonado, R.~Granda, K.~Chiluiza, C.~Conati, and
  S.~B. Shum.
\newblock Driving data storytelling from learning design.
\newblock In {\em Proceedings of the 8th International Conference on Learning
  Analytics and Knowledge}, pp. 131--140. ACM, 2018.

\bibitem{Fu2017}
X.~Fu, A.~Shimada, H.~Ogata, Y.~Taniguchi, and D.~Suehiro.
\newblock Real-time learning analytics for c programming language courses.
\newblock In {\em Proceedings of the Seventh International Learning Analytics
  \&\#38; Knowledge Conference}, LAK '17, pp. 280--288. ACM, New York, NY, USA,
  2017. doi: {{%
10\hspace{.1pt}\discretionary{.}{%
}{.}\hspace{.4pt}1145\discretionary{/}{%
}{/}3027385\hspace{.1pt}\discretionary{.}{%
}{.}\hspace{.4pt}3027407}}


\bibitem{govaerts12}
S.~Govaerts, K.~Verbert, E.~Duval, and A.~Pardo.
\newblock {The Student Activity Meter for Awareness and Self-reflection}.
\newblock In {\em CHI '12 Extended Abstracts on Human Factors in Computing
  Systems}, CHI EA '12, pp. 869--884. ACM, New York, NY, USA, 2012. doi: {{%
10\hspace{.1pt}\discretionary{.}{%
}{.}\hspace{.4pt}1145\discretionary{/}{%
}{/}2212776\hspace{.1pt}\discretionary{.}{%
}{.}\hspace{.4pt}2212860}}


\bibitem{govaerts10}
S.~Govaerts, K.~Verbert, J.~Klerkx, and E.~Duval.
\newblock {Visualizing Activities for Self-reflection and Awareness}.
\newblock In {\em International Conference on Web-based Learning}, pp. 91--100.
  Springer, 2010.

\bibitem{Grissom2003}
S.~Grissom, M.~F. McNally, and T.~Naps.
\newblock Algorithm visualization in cs education: Comparing levels of student
  engagement.
\newblock In {\em Proceedings of the 2003 ACM Symposium on Software
  Visualization}, SoftVis '03, pp. 87--94. ACM, New York, NY, USA, 2003. doi:
  {{%
10\hspace{.1pt}\discretionary{.}{%
}{.}\hspace{.4pt}1145\discretionary{/}{%
}{/}774833\hspace{.1pt}\discretionary{.}{%
}{.}\hspace{.4pt}774846}}


\bibitem{hattie06}
J.~A. Hattie, G.~T. Brown, L.~Ward, S.~E. Irving, and P.~J. Keegan.
\newblock {Formative Evaluation of an Educational Assessment Technology
  Innovation: Developers' Insights into Assessment Tools for Teaching and
  Learning (asTTle)}.
\newblock {\em Journal of Multi-Disciplinary Evaluation}, 5(3):1--54, 2006.

\bibitem{jacobs05}
K.~L. Jacobs.
\newblock {Investigation of Interactive Online Visual Tools for the Learning of
  Mathematics}.
\newblock {\em International Journal of Mathematical Education in Science and
  Technology}, 36(7):761--768, 2005. doi: {{%
10\hspace{.1pt}\discretionary{.}{%
}{.}\hspace{.4pt}1080\discretionary{/}{%
}{/}00207390500271149}}


\bibitem{Jivet2018}
I.~Jivet, M.~Scheffel, M.~Specht, and H.~Drachsler.
\newblock License to evaluate: Preparing learning analytics dashboards for
  educational practice.
\newblock In {\em Proceedings of the 8th International Conference on Learning
  Analytics and Knowledge}, LAK '18, pp. 31--40. ACM, New York, NY, USA, 2018.
  doi: {{%
10\hspace{.1pt}\discretionary{.}{%
}{.}\hspace{.4pt}1145\discretionary{/}{%
}{/}3170358\hspace{.1pt}\discretionary{.}{%
}{.}\hspace{.4pt}3170421}}


\bibitem{kont2017}
M.~Kont, M.~Pihelgas, K.~Maennel, B.~Blumbergs, and T.~Lepik.
\newblock Frankenstack: Toward real-time red team feedback.
\newblock In {\em Military Communications Conference (MILCOM), MILCOM 2017-2017
  IEEE}, pp. 400--405. IEEE, 2017.

\bibitem{kumar2011}
S.~Kumar, R.~Zafarani, and H.~Liu.
\newblock Understanding user migration patterns in social media.
\newblock In {\em Twenty-Fifth AAAI Conference on Artificial Intelligence},
  2011.

\bibitem{Leony2012}
D.~Leony, A.~Pardo, L.~de~la Fuente~Valent\'{\i}n, D.~S. de~Castro, and C.~D.
  Kloos.
\newblock Glass: A learning analytics visualization tool.
\newblock In {\em Proceedings of the 2Nd International Conference on Learning
  Analytics and Knowledge}, LAK '12, pp. 162--163. ACM, New York, NY, USA,
  2012. doi: {{%
10\hspace{.1pt}\discretionary{.}{%
}{.}\hspace{.4pt}1145\discretionary{/}{%
}{/}2330601\hspace{.1pt}\discretionary{.}{%
}{.}\hspace{.4pt}2330642}}


\bibitem{loh15}
C.~S. Loh, Y.~Sheng, and D.~Ifenthaler.
\newblock {\em {Serious Games Analytics: Methodologies for Performance
  Measurement, Assessment, and Improvement}}.
\newblock Advances in Game-Based Learning. Springer International Publishing,
  2015.

\bibitem{Matcha2019}
W.~Matcha, D.~Ga\v{s}evi\'{c}, N.~A. Uzir, J.~Jovanovi\'{c}, and A.~Pardo.
\newblock Analytics of learning strategies: Associations with academic
  performance and feedback.
\newblock In {\em Proceedings of the 9th International Conference on Learning
  Analytics \& Knowledge}, LAK19, pp. 461--470. ACM, New York, NY, USA, 2019.
  doi: {{%
10\hspace{.1pt}\discretionary{.}{%
}{.}\hspace{.4pt}1145\discretionary{/}{%
}{/}3303772\hspace{.1pt}\discretionary{.}{%
}{.}\hspace{.4pt}3303787}}


\bibitem{michael2005}
D.~R. Michael and S.~L. Chen.
\newblock {\em {Serious Games: Games that Educate, Train, and Inform}}.
\newblock Muska \& Lipman/Premier-Trade, 2005.

\bibitem{nagarajan12}
A.~Nagarajan, J.~M. Allbeck, A.~Sood, and T.~L. Janssen.
\newblock {Exploring Game Design for Cybersecurity Training}.
\newblock In {\em Cyber Technology in Automation, Control, and Intelligent
  Systems (CYBER), 2012 IEEE International Conference on}, pp. 256--262. IEEE,
  2012.

\bibitem{questionmark}
{Questionmark Computing Limited}.
\newblock Questionmark perception.
\newblock \url{https://www.questionmark.com}.
\newblock (Accessed on 2018-06-13).

\bibitem{Rubin1994}
J.~Rubin.
\newblock {\em Handbook of Usability Testing: How to Plan, Design, and Conduct
  Effective Tests}.
\newblock John Wiley \& Sons, Inc., New York, NY, USA, 1st ed., 1994.

\bibitem{netwars}
{SANS Institute}.
\newblock {NetWars: DFIR Tournament}.
\newblock (Accessed on 2018-06-13).

\bibitem{schwendimann2017}
B.~A. {Schwendimann}, M.~J. {Rodríguez-Triana}, A.~{Vozniuk}, L.~P. {Prieto},
  M.~S. {Boroujeni}, A.~{Holzer}, D.~{Gillet}, and P.~{Dillenbourg}.
\newblock Perceiving learning at a glance: A systematic literature review of
  learning dashboard research.
\newblock {\em IEEE Transactions on Learning Technologies}, 10(1):30--41, Jan
  2017. doi: {{%
10\hspace{.1pt}\discretionary{.}{%
}{.}\hspace{.4pt}1109\discretionary{/}{%
}{/}TLT\hspace{.1pt}\discretionary{.}{%
}{.}\hspace{.4pt}2016\hspace{.1pt}\discretionary{.}{%
}{.}\hspace{.4pt}2599522}}


\bibitem{shi2019}
Y.~Shi, Y.~Liu, H.~Tong, J.~He, G.~Yan, and N.~Cao.
\newblock Visual analytics of anomalous user behaviors: A survey.
\newblock {\em arXiv preprint arXiv:1905.06720}, 2019.

\bibitem{stewart09}
K.~E. Stewart, J.~W. Humphries, and T.~R. Andel.
\newblock {Developing a Virtualization Platform for Courses in Networking,
  Systems Administration and Cyber Security Education}.
\newblock In {\em Proceedings of the 2009 Spring Simulation Multiconference},
  p.~65. Society for Computer Simulation International, 2009.

\bibitem{thom2012}
D.~Thom, H.~Bosch, S.~Koch, M.~W{\"o}rner, and T.~Ertl.
\newblock Spatiotemporal anomaly detection through visual analysis of
  geolocated twitter messages.
\newblock In {\em 2012 IEEE Pacific Visualization Symposium}, pp. 41--48. IEEE,
  2012.

\bibitem{Vatrapu11}
R.~Vatrapu, C.~Teplovs, N.~Fujita, and S.~Bull.
\newblock {Towards Visual Analytics for Teachers' Dynamic Diagnostic
  Pedagogical Decision-making}.
\newblock In {\em Proceedings of the 1st International Conference on Learning
  Analytics and Knowledge}, LAK '11, pp. 93--98. ACM, New York, NY, USA, 2011.
  doi: {{%
10\hspace{.1pt}\discretionary{.}{%
}{.}\hspace{.4pt}1145\discretionary{/}{%
}{/}2090116\hspace{.1pt}\discretionary{.}{%
}{.}\hspace{.4pt}2090129}}


\bibitem{svabensky2018}
V.~\v{S}v\'{a}bensk\'{y}, J.~Vykopal, M.~Cermak, and M.~La\v{s}tovi\v{c}ka.
\newblock Enhancing cybersecurity skills by creating serious games.
\newblock In {\em Proceedings of the 23rd Annual ACM Conference on Innovation
  and Technology in Computer Science Education}, ITiCSE 2018, pp. 194--199.
  ACM, New York, NY, USA, 2018. doi: {{%
10\hspace{.1pt}\discretionary{.}{%
}{.}\hspace{.4pt}1145\discretionary{/}{%
}{/}3197091\hspace{.1pt}\discretionary{.}{%
}{.}\hspace{.4pt}3197123}}


\bibitem{ITICSE18kypolab}
V.~\v{S}v\'{a}bensk\'{y}, J.~Vykopal, M.~\v{c}ermak, and M.~La\v{s}tovi\v{c}ka.
\newblock {Enhancing Cybersecurity Skills by Creating Serious Games}.
\newblock In {\em Proceedings of the 23rd Annual ACM Conference on Innovation
  and Technology in Computer Science Education}, ITiCSE '18. ACM, 2018. doi:
  {{%
10\hspace{.1pt}\discretionary{.}{%
}{.}\hspace{.4pt}1145\discretionary{/}{%
}{/}3197091\hspace{.1pt}\discretionary{.}{%
}{.}\hspace{.4pt}3197123}}


\bibitem{vykopal2018}
J.~Vykopal, R.~O{\v{s}}lej{\v{s}}ek, K.~Bursk{\'a}, and
  K.~Z{\'a}kop{\v{c}}anov{\'a}.
\newblock Timely feedback in unstructured cybersecurity exercises.
\newblock In {\em Proceedings of the 49th ACM Technical Symposium on Computer
  Science Education}, pp. 173--178. ACM, 2018.

\bibitem{KYPO}
J.~Vykopal, R.~Oslejsek, P.~Celeda, M.~Vizvary, and D.~Tovarnak.
\newblock {KYPO Cyber Range: Design and Use Cases}.
\newblock In {\em Proceedings of the 12th International Conference on Software
  Technologies -- Volume 1: ICSOFT}, pp. 310--321. SciTePress, 2017. doi: {{%
10\hspace{.1pt}\discretionary{.}{%
}{.}\hspace{.4pt}5220\discretionary{/}{%
}{/}0006428203100321}}


\bibitem{weiss2017magazine}
R.~Weiss, F.~Turbak, J.~Mache, and M.~E. Locasto.
\newblock Cybersecurity education and assessment in edurange.
\newblock {\em IEEE Security \& Privacy}, 15(3):90--95, 2017.

\bibitem{Werther2011}
J.~Werther, M.~Zhivich, T.~Leek, and N.~Zeldovich.
\newblock {Experiences in Cyber Security Education: The MIT Lincoln Laboratory
  Capture-the-flag Exercise}.
\newblock In {\em Proceedings of the 4th Conference on Cyber Security
  Experimentation and Test}, CSET'11, pp. 12--12. USENIX Association, Berkeley,
  CA, USA, 2011.

\bibitem{Yang2010}
Z.~Yang, J.~Guo, K.~Cai, J.~Tang, J.~Li, L.~Zhang, and Z.~Su.
\newblock Understanding retweeting behaviors in social networks.
\newblock In {\em Proceedings of the 19th ACM International Conference on
  Information and Knowledge Management}, CIKM '10, pp. 1633--1636. ACM, New
  York, NY, USA, 2010. doi: {{%
10\hspace{.1pt}\discretionary{.}{%
}{.}\hspace{.4pt}1145\discretionary{/}{%
}{/}1871437\hspace{.1pt}\discretionary{.}{%
}{.}\hspace{.4pt}1871691}}


\bibitem{picoctf}
K.~Zhang, S.~Dong, G.~Zhu, D.~Corporon, T.~McMullan, and S.~Barrera.
\newblock picoctf 2013-toaster wars: When interactive storytelling game meets
  the largest computer security competition.
\newblock In {\em Games Innovation Conference (IGIC), 2013 IEEE International},
  pp. 293--299. IEEE, 2013.

\end{thebibliography}

\end{document}